\documentclass[aip,graphicx]{revtex4-1}
\usepackage{graphicx}
\usepackage{amsmath}
\usepackage{multirow}
\usepackage{hyperref}
\usepackage[finalnew]{trackchanges}
\usepackage[justification=centerlast,font=scriptsize]{caption}
\draft 

\usepackage{xcolor}

\hypersetup{
    colorlinks=true,
    linkcolor=blue,
    filecolor=magenta,      
    urlcolor=cyan,
    pdftitle={Overleaf Example},
    pdfpagemode=FullScreen,
    }

\begin{document}


\title{Nonlinear Landau resonant interaction between whistler waves and electrons: Excitation of electron acoustic waves} 



\author{Donglai Ma}
\email[]{dma96@atmos.ucla.edu}
\affiliation{Department of Atmospheric and Oceanic Sciences, University of California, Los Angeles, CA, 90095, USA.}

\author{Xin An}
\affiliation{Department of Earth, Space and Planetary Sciences, University of California, Los Angeles, CA, 90095, USA.}

\author{Anton Artemyev}
\affiliation{Department of Earth, Space and Planetary Sciences, University of California, Los Angeles, CA, 90095, USA.}

\author{Jacob Bortnik}
\affiliation{Department of Atmospheric and Oceanic Sciences, University of California, Los Angeles, CA, 90095, USA.}

\author{Vassilis Angelopoulos}
\affiliation{Department of Earth, Space and Planetary Sciences, University of California, Los Angeles, CA, 90095, USA.}

\author{Xiao-Jia Zhang}
\affiliation{Department of Physics, University of Texas at Dallas, Richardson, TX, 75080, USA.}
\affiliation{Department of Earth, Space and Planetary Sciences, University of California, Los Angeles, CA, 90095, USA.}


\date{\today}

\begin{abstract}
Electron acoustic waves (EAWs), as well as electron-acoustic solitary structures, play a crucial role in thermalization and acceleration of electron populations in Earth's magnetosphere. These waves are often observed in association with whistler-mode waves, but the detailed mechanism of EAW and whistler wave coupling is not yet revealed. We investigate the excitation mechanism of EAWs and their potential relation to whistler waves using particle-in-cell simulations. Whistler waves are first excited by electrons with a temperature anisotropy perpendicular to the background magnetic field. Electrons trapped by these whistler waves through nonlinear Landau resonance form localized field-aligned beams, which subsequently excite EAWs. By comparing the growth rate of EAWs and the phase mixing rate of trapped electron beams, we obtain the critical condition for EAW excitation, which is consistent with our simulation results across a wide region in parameter space. These results are expected to be useful in the interpretation of concurrent observations of whistler-mode waves and nonlinear solitary structures, and may also have important implications for investigation of cross-scale energy transfer in the near-Earth space environment.
\end{abstract}


\maketitle 

\section{Introduction}




During geomagnetically active times, fast plasma flows in the Earth's plasma sheet transport energetic particles into the inner magnetosphere and form injection fronts (or dipolarization fronts; see reviews in \onlinecite{birn2012particle,gabrielse2023mesoscale,birn2021magnetotail}). In the leading edge of injection fronts, the magnetic $B_z$ component (in the Geocentric Solar Magnetospheric coordinate) typically has an abrupt enhancement, indicating a ``dipolarization'' of the geomagnetic field\cite{baker1996neutral,liu2016dipolarizing,angelopoulos2008tail,runov2009themis}. Injections consist of abruptly enhanced fluxes of high-energy ions and electrons in the energy range of $10$s -- $100$s of keV, which provide free energy to various plasma waves\cite{chaston2012energy,chaston2014observations,malaspina2018census}. Indeed, a broad spectrum of electromagnetic emissions, extending from Doppler-shifted kinetic Alfv\'en waves of a few Hz to electron cyclotron harmonic waves of $\sim 10$\,kHz, is embedded within dipolarization fronts and constitutes a significant fraction of the total energy transport in fast plasma flows. Among these emissions, whistler waves are excited by injected energetic electrons with a perpendicular temperature anisotropy \cite{hwang2007statistical,le2009quasi,breuillard2016multispacecraft,zhang2019energy,li2009evaluation}. Electron acoustic waves (EAWs), as well as electron-acoustic solitary structures, identified as broadband electrostatic turbulence, are often observed in association with whistler waves\cite{reinleitner1982chorus,mozer2015time,li2017chorus,chen2022high,agapitov2018nonlinear,vasko2018electrostatic,dillard2018electron,osmane2017subcritical}. Under typical conditions of plasma injections into the inner magnetosphere, EAWs and their related solitary structures effectively scatter electrons of $10$\,eV--$1$\,keV in both pitch angle and energy\cite{osmane2014threshold,artemyev2014thermal,vasko2017diffusive,shen2020potential}, whereas whistler waves provide electron scattering in the $1$--$100$s\,keV energy range\cite{summers1998relativistic,horne1998potential,horne2003relativistic,shprits2008review,thorne2013rapid,reeves2013electron}. The concurrence of EAWs and whistler waves potentially results in a wide energy range for electron precipitation and acceleration\cite{ma2016strong}.

Linear wave theory, confirmed by spacecraft observations, indicates that even slightly oblique whistler waves (e.g., $\lesssim 30^\circ$ of wave normal angle with respect to the background magnetic field) have finite parallel electric fields, and trap electrons through the nonlinear Landau resonance \cite{agapitov2015nonlinear,li2017chorus,an2019unified,zhang2022superfast}. It was demonstrated that such trapped electron populations can form beams, that are unstable to the generation of EAWs or other electron-acoustic solitary structures with spatial scales on the order of tens of Debye lengths (e.g., double layers, phase space holes, also known as time domain structures; see Refs. \onlinecite{mozer2015time,malaspina2018census,vasko2017electron}). The ratio of Landau resonant velocity to electron thermal velocity controls the type of nonlinear wave structures generated by electron populations trapped by whistler waves \cite{an2019unified}. The same mechanism of electron Landau trapping by lower frequency waves and further generation of higher frequency electrostatic structures has been confirmed to work for the excitation of electron-acoustic solitary structures through interactions between kinetic Alfv\'en waves and thermal electrons \cite{an2021nonlinear}. It is worth mentioning that an alternative mechanism was proposed for the formation of electric field spikes, such as nonlinear fluid steepening of electron acoustic modes\cite{vasko2018electrostatic,agapitov2018nonlinear}. The link between whistler waves (or kinetic Alfv\'en waves) and EAWs (or solitary structures) provides a potentially important channel of energy transfer: Injected ions and electrons accelerated by the electromagnetic fields of dipolarization fronts at the macroscale (tens to hundreds of ion inertial length) first excite kinetic Alfv\'en and whistler waves at the intermediate scale (a few ion or electron inertial lengths), which subsequently generate EAWs and solitary structures at the microscale (tens of Debye lengths) and eventually deposit energy into thermal electrons\cite{an2021nonlinear,vasko2015thermal}. Such energy transfer from the macroscale to the microscale may contribute to electron thermalization and heating during one of the most energetic processes in the Earth's magnetosphere - the plasma injection and braking of fast plasma flows in the inner magnetosphere\cite{angelopoulos2002plasma,stawarz2015generation,ergun2015large}.

It is the aim of this study to explore the coupling from whistler waves to EAWs using particle-in-cell (PIC) simulations and to determine the favorable conditions for such coupling to occur. It is organized as follows: In Section \ref{sec:setup}, we briefly describe the computational setup, in which whistler waves are naturally generated by energetic electrons with a perpendicular temperature anisotropy. In Section \ref{sec:excitation}, we investigate how trapped electron beams are formed through nonlinear Landau resonance between electrons and whistler waves, and how such electron beams excite EAWs and make EAWs survive. In Section \ref{sec:condition}, we derive the critical condition for EAW excitation and confirm its validity by comparing it with simulation results. We summarize our results in Section \ref{sec:summary}.

\section{Computational setup\label{sec:setup}}

We use a fully relativistic, electromagnetic PIC code called OSIRIS 4.0 \cite{fonseca2002osiris}. Our simulations have two dimensions (2D) in configuration space and three dimensions in velocity space. The computation domain in the $x$-$y$ plane consists of $625 \times 625$ cells. Each cell contains 400 particles. We use periodic boundary conditions for both particles and fields. The background magnetic field $\mathbf{B}_0$ is in the $x$ direction. The normalized strength of $\mathbf{B}_0$ is set as ${\omega_{ce}}/{\omega_{pe}} = 0.25$, typical in the generation region of whistler waves in the inner magnetosphere\cite{fu2014whistler,tao2011evolution}. Here $\omega_{ce}$ is the electron gyrofrequency, and $\omega_{pe}$ is the electron plasma frequency. The plasma is initially uniform in space. Because our frequency range of interest is $\omega \gg \omega_{ci} \text{ (the ion gyrofrequency)}$, ions are immobile as a charge-neutralizing background. Electrons are initialized with a single bi-maxwellian distribution with a temperature anisotropy $A = T_{\perp} / T_{\parallel} > 1$. \add{The bi-maxwellian model is a theoretical and typical construct in order to carry out the simulations. It is worth noting that the observed distributions sometimes may deviate significantly from Maxwellians}\cite{fu2014whistler}. The electron parallel beta is defined as
\begin{equation}
\beta_{\|} = \frac{n_0 m_e v_{T_{\parallel}}^2}{B_0^2 / 8 \pi} = \frac{2 v_{T_{\parallel}}^{2}}{\left(c \omega_{ce} / \omega_{pe}\right)^{2}} = \frac{2 v_{T_{\parallel}}^{2}}{v_{Ae}^2} ,
\end{equation}
where $m_e$ is the electron mass, $c$ is the speed of light, $n_0$ is the plasma density, $v_{Ae}$ is the electron Alfv\'en speed, and $v_{T_{\parallel}}$ is the electron thermal velocity in the parallel direction. $\beta_{\parallel}$ describes the magnitude of the electron thermal velocity relative to the characteristic whistler phase velocity ($0.5 v_{Ae}$ being the whistler phase velocity at $0.5 \omega_{ce}$ and 0 degree wave normal angle). Given $A$ and $\beta_{\parallel}$, we determine the initial $T_{\perp}$ and $T_{\parallel}$, and initialize the electron velocity distribution.
The cell length $\Delta x$ in both directions is set between $\lambda_D$ and $2 \lambda_D$ where the initial electron Debye length $\lambda_D  = v_{T_{\parallel}}/\omega_{pe}$ (neglecting ions). For 2D simulations, the time step $\Delta t$ is constrained by the Courant condition:
\begin{equation}
    \Delta t \lesssim \frac{\Delta x}{\sqrt{2} c } .
\end{equation}
To understand the critical condition of EAW excitation, we scan the parameter space of $\beta_{\parallel}$ and $T_{\perp} / T_{\parallel}$. In this scan, $\beta_{\parallel}$ is varied from $0.005$ to $0.3$ with a logarithmic step, and $T_{\perp} / T_{\parallel}$ is varied in the sequence $(3.5, 4, 4.5, 5)$. The detailed parameters in each simulation are shown in Table \ref{tab:my-table}. \add{Such choice is made to facilitate numerical work and the observed anisotropies in space may deviate from the assumed values.}\cite{an2017parameter}.

\begin{table}[tphb]
\centering
\resizebox{0.8\textwidth}{!}{%
\begin{tabular}{|c|c|c|c|c|c|c|c|c|}
\cline{1-4} \cline{6-9}
case No. & $\beta_{\parallel}$ & $\Delta x$ & $\Delta t$ &  & case No. & $\beta_{\parallel}$ & $\Delta x$ & $\Delta t$ \\ \cline{1-4} \cline{6-9} 
1 -  4  & 0.00500 & 0.0216 & 0.0145 &  & 17 - 20 & 0.05268 & 0.040 & 0.0270 \\ \cline{1-4} \cline{6-9} 
5 -  8  & 0.00901 & 0.0216 & 0.0145 &  & 21 - 24 & 0.09491 & 0.056 & 0.0370 \\ \cline{1-4} \cline{6-9} 
9 - 12  & 0.01623 & 0.0216 & 0.0145 &  & 25 - 28 & 0.17100 & 0.072 & 0.0476 \\ \cline{1-4} \cline{6-9} 
13 - 16 & 0.02924 & 0.0300 & 0.0200 &  & 29 - 32 & 0.30808 & 0.098 & 0.0670 \\ \cline{1-4} \cline{6-9} 
\end{tabular}%
}
\caption{The detailed parameters for $32$ simulations. The unit of cell length $\Delta x$ is $c/\omega_{pe}$, the unit of time step $\Delta t$ is $\omega_{pe}^{-1}$. Each group has four cases, corresponding to temperature anisotropies $T_{\perp} / T_{\parallel} = 3.5, 4, 4.5 \text{ and } 5$.}
\label{tab:my-table}
\end{table}

\section{Excitation of EAWs by whistler waves through Nonlinear Electron Trapping}\label{sec:excitation}

Whistler waves can be excited by the free energy provided by electron perpendicular temperature anisotropy\cite{kennel1966low}. Linear kinetic theory and PIC simulations in previous works \cite{gary2011whistler,yue2016relationship,an2017parameter} have shown that the dominant mode of maximum growth rate is in the parallel direction for $\beta_{\parallel  }\gtrsim0.025$ and shifts to the oblique direction for $\beta_{\parallel}\lesssim 0.025$. Here we examine the evolution of the fields, as well as the electron Landau trapping in these fields, in the two different $\beta_{\parallel}$ regimes, and demonstrate that both regimes support electron acoustic wave (EAW) excitation.

\subsection{Small $\beta_{\parallel}$ regime}

\begin{figure}[tphb]
\noindent\includegraphics[width=\textwidth]{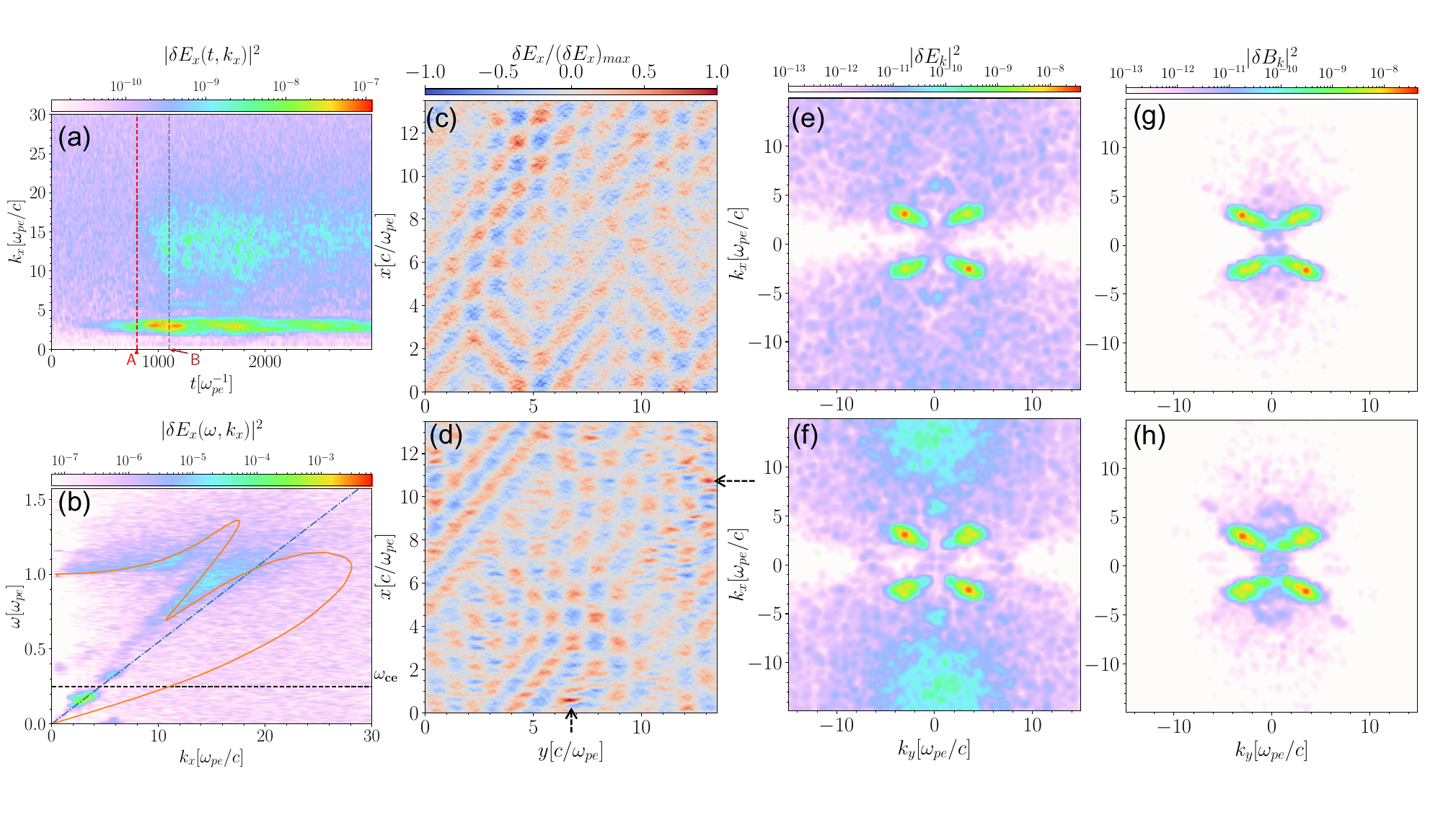}%
\caption{\label{Figure 1} Whistler anisotropy instability and the associated excitation of EAWs for $\beta_{\parallel} = 0.0091$ and $T_{\perp} / T_{\|}=5.0$. (a) The wave number spectrum of the parallel electric field $\delta E_x$ as a function of time. (b) The dispersion diagram of $\delta E_x$ as a function of frequency and wave number. The black dashed line represents the electron cyclotron frequency $\omega_{ce} = 0.25 \omega_{pe}$. The blue dash-dotted line is the parallel phase velocity of the whistler wave $v_{phase, \parallel} = 0.0546c = 3.25 v_{T,\parallel} $. The orange line shows the EAW dispersion calculated from the electron distribution at $ t = 1100 \omega_{pe}^{-1}$. (c)-(d) The pattern of $\delta E_x$ at two time snapshots, $ t = 800 \omega_{pe}^{-1}$ and $ t = 1100 \omega_{pe}^{-1}$, respectively. (e)-(f) The power density of $\delta E_x$ in the wave number space at $ t = 800 \omega_{pe}^{-1}$ and $ t = 1100 \omega_{pe}^{-1}$. (g)-(h) The power density of $\delta B_y$ in the wave number space at $ t = 800 \omega_{pe}^{-1}$ and $ t = 1100 \omega_{pe}^{-1}$, respectively.}  %
\end{figure}

Figure \ref{Figure 1} illustrates the wave characteristics in the small $\beta_{\parallel}$ regime of $\beta_{\parallel} = 0.0091$ and $T_{\perp} / T_{\|}=5.0$ (case 8 in Table \ref{tab:my-table}). The whistler waves are first excited at $t \sim 300 \omega_{pe}^{-1}$ at the parallel wave number $k_x \sim 3 {\omega_{pe}}/{c}$. We denote the parallel whislter wave number as $k_\parallel^{\prime}$ and the EAW wave number as $k_\parallel$. The EAWs start to be excited at $t \sim 1000 \omega_{pe}^{-1}$ with the {wave number} $k_\parallel$ ranging from  $5 \omega_{pe}/c$ to $20 \omega_{pe}/c$ [Figure \ref{Figure 1}(a)]. The upper band whistler-mode with maximum power, located at $\omega = 0.18 \omega_{pe} = 0.72 \omega_{ce}$ and $k_{\parallel}^{\prime} = 3.26 \omega_{pe}/c$, has a parallel phase speed $v_{ph,\parallel} = \omega / k_{\parallel}^{\prime} = 0.055c $, which is about the same as (or slightly smaller than) that of EAWs [Figure \ref{Figure 1}(b)]. Two representative time snapshots of wave fields (including $\delta E_x$ and $\delta B_y$) at $t_A = 800 \omega_{pe}^{-1}$ and $t_B = 1100 \omega_{pe}^{-1}$, before and after the excitation of EAWs, respectively, are displayed in Figures \ref{Figure 1}(c)-(h). The electromagnetic, relatively long-wavelength ($\lambda \sim 1.7 c/\omega_{pe}$) whistler waves propagate in the oblique direction with a wave normal angle (WNA) $\sim 45^{\circ}$, whereas the electrostatic, short-wavelength EAWs ($\lambda \sim 0.5 c /\omega_{pe} \sim 30 \lambda_{De}$) propagate in the parallel direction. Such electron-acoustic mode manifests as solitary structures in certain spatial domains [see such domains pointed by arrows in Figure \ref{Figure 1}(d)] and hence has a broadband wave number spectrum.

The perturbed magnetic field amplitude reaches $\sim 0.01 B_0$ at $t_B = 1100 \omega_{pe}^{-1}$. For large-amplitude whistler waves propagating in oblique directions, their parallel electric field can trap the electrons moving near the parallel phase speed $v_{ph,\parallel}$ (i.e., the Landau resonant velocity) in its potential well, which is the so-called nonlinear Landau resonance \cite{o1965collisionless}. The response of trapped electrons to whistler waves is characterized by the formation of electron beams in the resonant islands around $\pm v_{ph,\parallel}$ as shown in Figure \ref{Figure 3}(b). The resonant electrons are accelerated in the phase of $v_x \cdot \delta E_x < 0$, whereas they are decelerated in the phase of $v_x \cdot \delta E_x > 0$.  This transport process in the velocity space gives rise to spatially modulated beams, which subsequently excite time domain structures (TDSs) as shown in Figure \ref{Figure 1}(d). These TDSs are identified as the nonlinear electron-acoustic mode \cite{valentini2006excitation,an2021nonlinear} and survive Landau damping because of the plateau distribution created by the electron trapping [Figure \ref{Figure 4}(a)]. It is worth noting that the beam velocity is slightly larger than $v_{ph,\parallel}$ [Figure \ref{Figure 3}(b)], which makes the EAW phase velocity slightly larger than $v_{ph,\parallel}$ located at the center of the resonant island [Figure \ref{Figure 1}(b)]. 

We further use the reduced, parallel velocity distribution obtained at $t_B = 1100 \omega_{pe}^{-1}$ [Figure \ref{Figure 4}] to numerically solve for the dispersion relation of EAWs. The plateau at the wave phase velocity (i.e., $\partial f_0 /\left.\partial v\right|_{\omega / k} \simeq 0$) allows retaining only the principal part of the integral around the Landau contour\cite{valentini2006excitation}:
\begin{equation}
\begin{gathered}
1-\frac{1}{k^2} \int_L d v \frac{\partial f_0 / \partial v}{v-\omega / k}=0 
\end{gathered}
\end{equation}
with
\begin{equation}
    \int_L d v \frac{\partial f_0 / \partial v}{v-\omega / k}=\mathrm{p.v.} \int_{-\infty}^{+\infty} d v \frac{\partial f_0 / \partial v}{v-\omega / k} ,
\end{equation}
where the subscript ``L'' denotes the Landau contour, and ``p.v.'' represents the principal value integral. The roots of the dispersion relation yield the orange solid curve in Figure \ref{Figure 1}(b). The ``thumb dispersion", computed from a Maxwellian distribution assuming an infinitesimal resonant island \cite{valentini2006excitation}, is modified to branch at the beam velocity (i.e., the parallel phase velocity of whistler waves) due to the finite width of the resonant island in our simulation. This calculation clearly shows that the EAWs are excited by trapped beams approximately at the parallel phase velocity of whistler waves.

\subsection{Large $\beta_{\parallel}$ regime}

\begin{figure}[tphb]
\noindent\includegraphics[width=\textwidth]{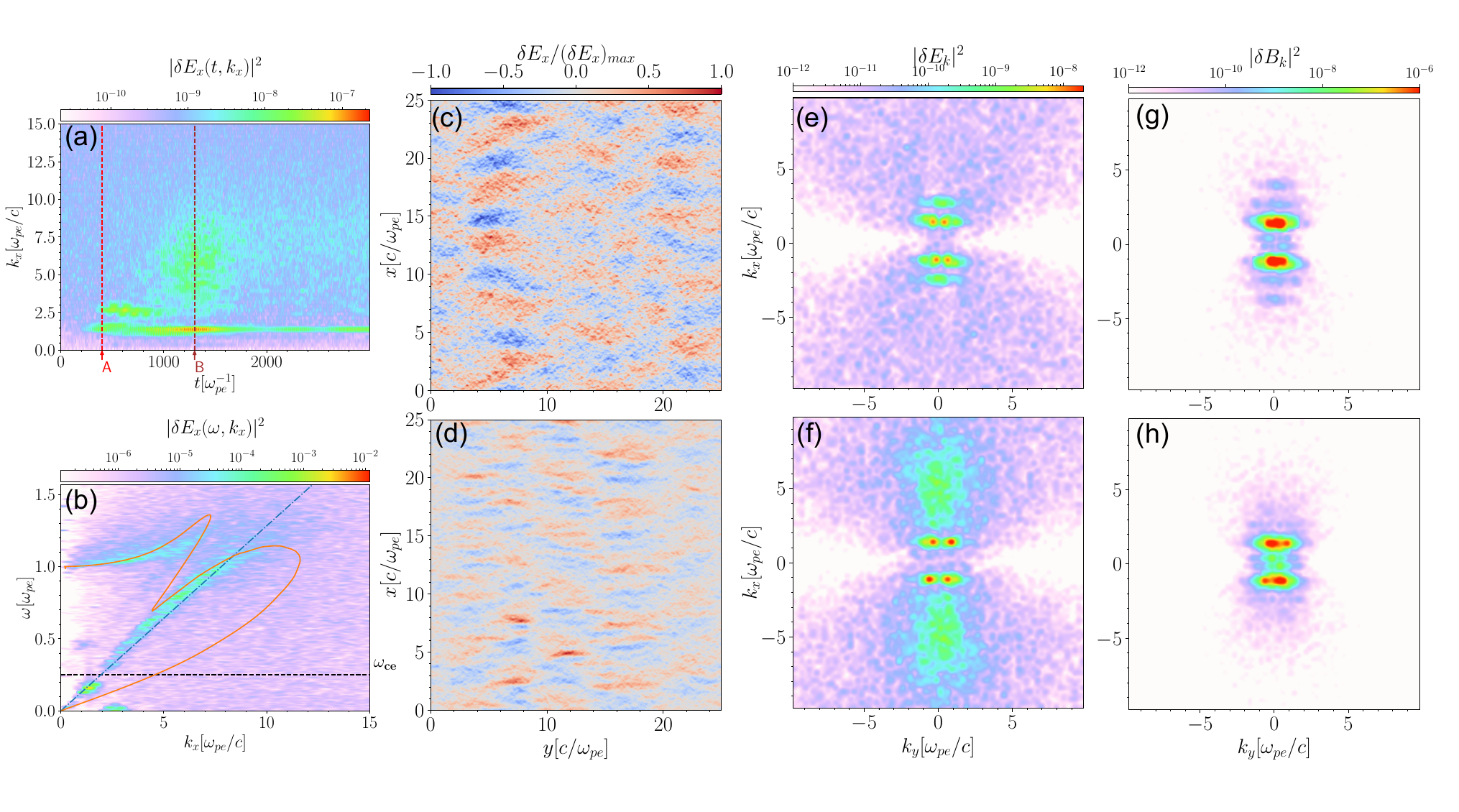}%
\caption{\label{Figure 2} The whistler anisotropy instability and associated excitation of EAWs for $\beta_{\parallel} = 0.05268$ and $T_{\perp} / T_{\|}=4.5$. The format is the same as that of Figure 1. The two selected snapshots are at $t_A = 400 \omega_{pe}^{-1}$ and $t_B = 1300 \omega_{pe}^{-1}$. The phase velocity of whistler waves is $v_{phase, \parallel} = 0.1283c = 3.16 v_{T,\parallel} $.}%
\end{figure}

Figure \ref{Figure 2} shows the wave characteristics for $\beta_{\parallel} = 0.05268$ and $T_{\perp} / T_{\| }=4.5$ (case 19 in Table \ref{tab:my-table}). The whistler mode with the maximum linear growth rate in this case is in the parallel direction. The excited upper band whistler mode waves with maximum power are located at $\omega = 0.16 \omega_{pe} = 0.64 \omega_{ce}$, and the energy density of the wave magnetic field is maximized at the wave number $k_{x,\max} = \pm 1.26 \omega_{pe} /c$ and $k_{y, \max} = 0$ [(Figures \ref{Figure 2}(g) and \ref{Figure 2}(h)].  It is interesting that there are low-frequency {quasi-electrostatic} modes at $\omega \sim 0$ and $k_x = 2 k_{x,\max}$ as shown in Figures \ref{Figure 2}(a), \ref{Figure 2}(b) and \ref{Figure 2}(e). \change{This standing electric field is generated because the Lorentz force is counter-propagating relative to the whistler waves, acting as a ponderomotive force, compressing and rarefying the electrons around the antinodes and nodes, respectively.}{This standing electric field is generated because counter-propagating whistler waves are naturally excited by electron temperature anisotropy in a uniform magnetic field, similar to the spacecraft observations of chorus source region in the equatorial magnetosphere }\cite{taubenschuss2015different}. \add{Specifically, the electron fluid should be force free parallel to the background magnetic field, $E_x + (\mathbf{v}_e \times \mathbf{B})_x = 0$ (neglecting density and temperature gradients for now). It can be shown that $(\mathbf{v}_e \times \mathbf{B})_x$ is finite by averaging over the fast wave periods for two counter-propagating whistler waves }\cite{sano2019ultrafast}. \add{Thus a standing electric field is generated.} 

\begin{figure}[tphb]
\noindent\includegraphics[width=\textwidth]{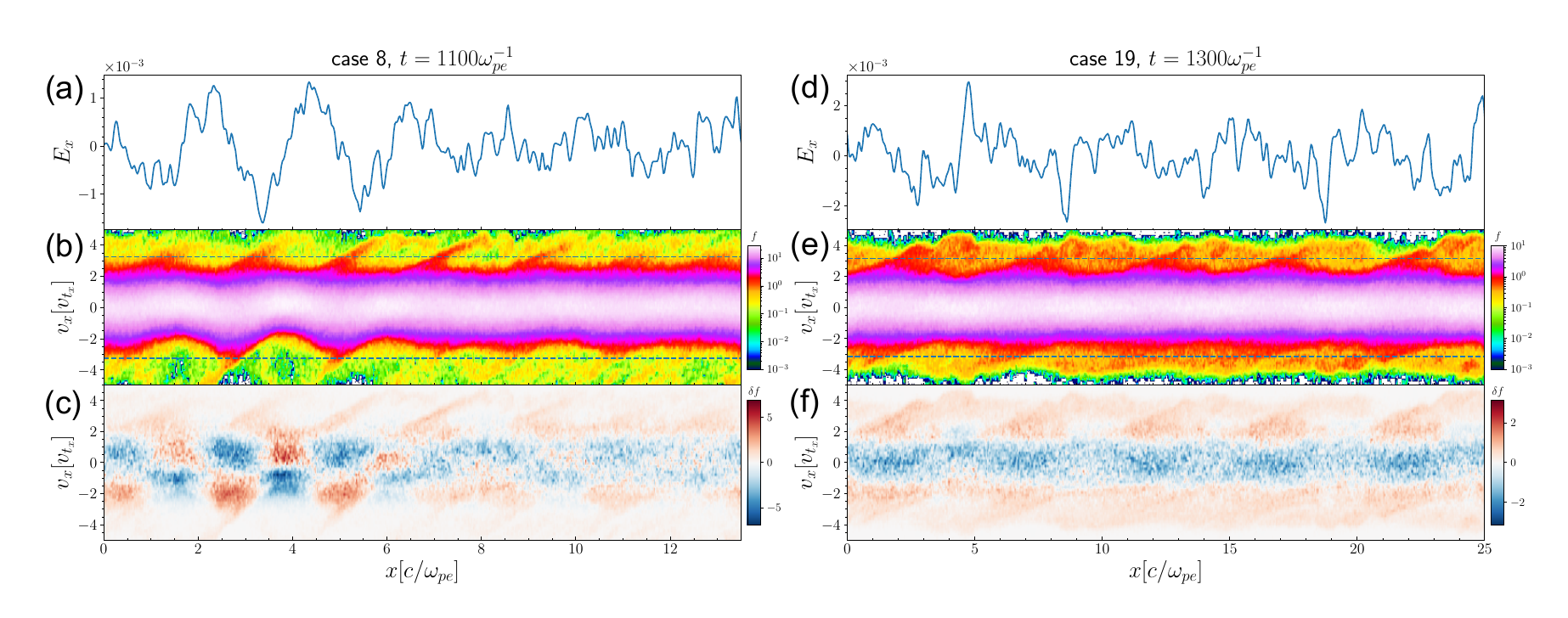}%
\caption{\label{Figure 3} Parallel electric fields and phase space densities in the two $\beta_{\parallel}$ regimes. Left column: Case 8 in the small $\beta_{\parallel}$ regime at $t = 1100 \omega_{pe}^{-1}$. (a) The parallel electric field as a function of position $x$. (b) The phase space density of electrons in $v_x$ -- $x$ space. The blue dashed line shows the parallel phase speed of whistler waves. (c) The difference of electron phase space densities between $t = 1100 \omega_{pe}^{-1}$ and $t = 0 \omega_{pe}^{-1}$. Right column: Case 19 in the large $\beta_{\parallel}$ regime at $t = 1300 \omega_{pe}^{-1}$. (d), (e) and (f) have the same format as (a), (b), and (c), respectively.}%
\end{figure}

\begin{figure}[htpb]
\noindent\includegraphics[width=\textwidth]{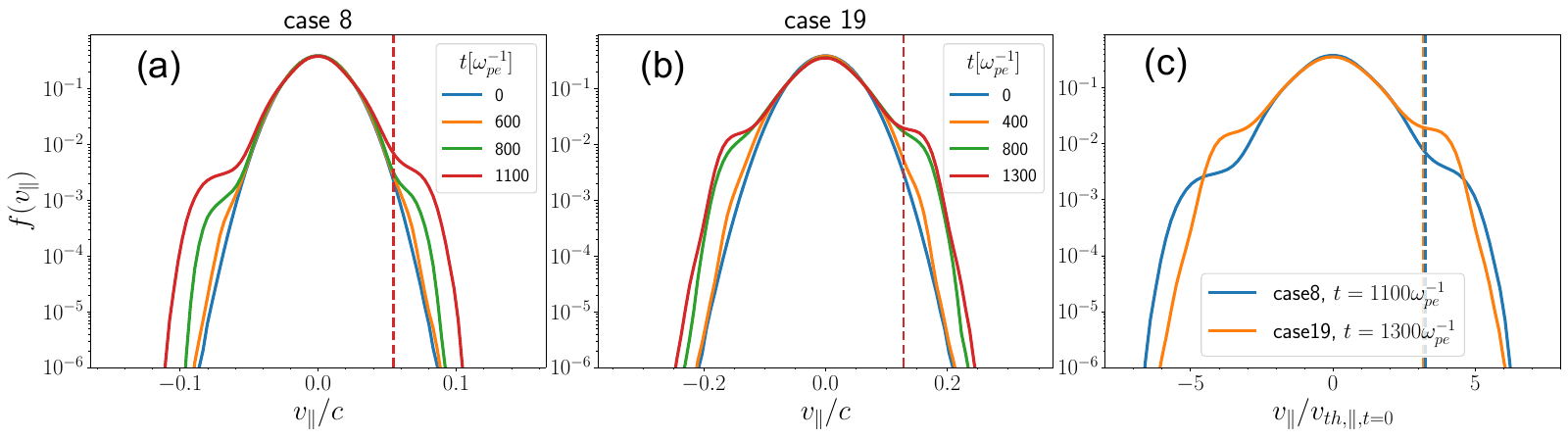}%
\caption{\label{Figure 4} The reduced electron distributions as a function of the parallel velocity for cases $8$ and $19$. (a) case 8 at $t = 0, 600, 800, \text{ and } 1100 \omega_{pe}^{-1}$. (b) case 19 at $t = 0, 400, 800, \text{ and } 1300 \omega_{pe}^{-1}$. (c) A comparison of phase space densities from both cases, for which velocities are normalized to the initial thermal velocity. The red dashed lines are the parallel phase speed of whistler waves}%
\end{figure}

The energy density of the slightly oblique whistler waves is still finite in the large $\beta_{\parallel}$ regime [Figures \ref{Figure 2}(c)--(h)]. The parallel electric field of these oblique waves excites electron-acoustic modes via the same mechanism as the small $\beta_{\parallel}$ regime. Figures \ref{Figure 2}(a) and \ref{Figure 2}(b) show that the electron-acoustic modes start to be excited at $t \sim 1000 \omega_{pe} ^{-1}$, and propagate at a phase velocity slightly larger than that of the whistler waves. These electron-acoustic modes appear as unipolar structures [Figure \ref{Figure 3}(d)], rather than wave-like structures in case $8$ [Figure \ref{Figure 3}(a)]. Such a difference is likely due to the larger beam density of case $19$ than that of case $8$ [Figures \ref{Figure 3}(b), \ref{Figure 3}(e), and \ref{Figure 4}], consistent with a previous study\cite{an2019unified}. 

\section{Critical Condition for EAW Excitation}\label{sec:condition}

The simulations demonstrate that EAWs can be excited in both low and high beta regimes. This leads to a question regarding the essential criteria for EAW excitation. As $\beta_{\parallel}$ changes, the parallel electric field amplitude and the phase velocity of the whistler waves change accordingly. These factors control the density and velocity of the trapped electron beams, which further control the growth rate of the beam instability. Furthermore, EAWs are subject to Landau damping, which depends on their phase velocities (or equivalently, the whistler wave phase velocities, we use same $v_{ph, \parallel}$ in the following). Moreover, trapped electrons undergo phase mixing\cite{o1965collisionless}, which smooths the beam distribution over time. The interplay between these processes determines the critical conditions for exciting EAWs driven by whistler waves. Our approach for solving this problem follows a similar method as outlined in Ref.~\onlinecite{an2021nonlinear}, which explored the interaction between kinetic Alfv\'en waves and thermal electrons.

The concept of trapped beams involves electrons being confined within the wave potential well, oscillating at a frequency $\omega_{t r}=\sqrt{e k_\parallel^{\prime}\left|\delta E_\parallel\right| / m_e}$ at the bottom of the potential well\cite{palmadesso1972resonance}. Consequently, the half-width of the trapping island or the plateau created by Landau resonance can be estimated as :
\begin{equation}\label{eq:v_tr}
\frac{\Delta v_{tr}}{v_{T_\parallel}} = \frac{2 \omega_{tr}}{k_\parallel^{\prime} v_{T_\parallel}} =\frac{2}{k_{\parallel }^{\prime} v_{T_\parallel}} \sqrt{\frac{e k_{\parallel }^{\prime} \cdot k_{\parallel }^{\prime} \delta \phi}{m_e}}=2 \sqrt{\frac{e \delta \phi}{T_\parallel}}.
\end{equation}
To compute the linear growth rate of the trapped electron beam, we model the parallel electron distribution by separating it into a background Maxwellian distribution $f_{0 e}=\frac{1}{\sqrt{2 \pi} v_{T \parallel}} \exp \left(-\frac{v_{\|}^{2}}{2 v_{T \parallel}^{2}}\right)$ (i.e., a reduced distribution normalized to a density of $1$) and a perturbed distribution $\delta f_{tr}$ around the parallel phase speed of the whistler wave $v_{ph,\parallel}$:
\begin{equation}\label{eq:f_tr}
    \delta f_{t r}=\Delta f \cdot \sin \left(\pi \frac{v_{\|}-v_{ph,\parallel}}{\Delta v_{t r}}\right)
\end{equation}
The magnitude of the perturbed distribution can be estimated as:
\begin{equation}\label{eq:Delta_f}
  \begin{gathered}
    \Delta f=\frac{f_{0 e}\left(v_{ph,\parallel}-\Delta v_{t r} / 2\right)-f_{0 e}\left(v_{ph,\parallel}+\Delta v_{t r} / 2\right)}{2}\\
    =\sum_{n=0}^{\infty} \frac{H_{2 n+1}\left(\frac{v_{ph,\parallel}}{\sqrt{2} v_{T \parallel}}\right)}{(2 n+1) !}\left(\frac{\Delta v_{t r}}{2 \sqrt{2} v_{T \parallel}}\right)^{2 n+1} f_{0 e}\left(v_{ph,\parallel}\right) ,  
  \end{gathered}
\end{equation}
where we perform a Taylor expansion at the whistler phase velocity $v_\parallel = v_{ph,\parallel}$, and the Hermite Polynomials $H_{2n + 1}(\cdot)$ are used for calculating derivatives of the Maxwellian distribution\cite{weber2003essential}. The growth rate of the beam instability\cite{o1968transition} can finally be written as:
\begin{equation}\label{eq:gamma-tr}
  \begin{gathered}
\frac{\gamma_{t r}}{k_{\|} v_{T \parallel}} =\frac{\pi}{2} \frac{\omega}{k_{\|} v_{T \parallel}} \frac{\omega_{p e}^{2}}{k_{\|}^{2}}\left(\frac{\partial}{\partial v_{\|}} \delta f_{t r}\right)_{v_{\|}=\omega / k_{\|}}  \sim \frac{\pi}{2} \frac{\omega}{k_{\|} v_{T \parallel}} \frac{\omega_{p e}^{2}}{k_{\|}^{2}}\left(\frac{\partial}{\partial v_{\|}} \delta f_{t r}\right)_{v_{\|}=v_{ph,\parallel}} \\
=\frac{\pi}{2} \frac{\omega}{k_{\|} v_{T \parallel}} \frac{1}{k_{\|}^{2} \lambda_{D}^{2}} \frac{\pi}{2 \sqrt{2}} v_{T \parallel} f_{0 e}\left(v_{ph,\parallel}\right) \times \sum_{n=0}^{\infty} \frac{H_{2 n+1}\left(\frac{v_{ph,\parallel}}{\sqrt{2} v_{T \parallel}}\right)\left(\frac{e \delta \phi}{2 T_{\parallel}}\right)^{n}}{(2 n+1) !} ,
 \end{gathered}
\end{equation}
where $\omega/ k _{\parallel}$ is the phase speed of EAWs, and the amplitude of electrostatic potential associated with the whistler wave is $\delta \phi =  \delta E_x / k_x$.
It is worth noting that the growth rate of trapped beam instabilities here is a function of $v_{ph,\parallel}/v_{T_\parallel}$ and $e \delta \phi / T_{\parallel}$ for a given $k_{\parallel}\lambda_D$. The zero-order term ($n = 0$) has no dependence on the wave amplitude, and it provides a positive growth rate due to the property $H_1(v_{ph,\parallel} / \sqrt{2} v_{T \parallel} ) > 0 $ for $v_{ph,\parallel} > 0$. The next higher order term ($n = 1$) is linearly proportional to the wave amplitude $\delta \phi$. The coefficient of this term $ H_3 (v_{ph,\parallel} / \sqrt{2} v_{T \parallel})  = 2 \sqrt{2} (v_{ph,\parallel}/v_{T \parallel}) \left((v_{ph,\parallel}/v_{T \parallel})^2 - 3 \right)$ indicates that the wave growth rate increases with $\delta \phi$ for $v_{ph,\parallel}/v_{T \parallel} > \sqrt{3}$, and decreases with $\delta \phi$ for $v_{ph,\parallel}/v_{T \parallel} < \sqrt{3}$.

On the other hand, the Landau damping rate of EAWs by the background distribution $f_{0e}$ can be written as:
\begin{equation}\label{eq:gamma_Le}
\begin{aligned}
\frac{\gamma_{L e}}{k_{\|} v_{T \parallel}} &=\frac{\pi}{2} \frac{\omega}{k_{\|} v_{T \parallel}} \frac{\omega_{p e}^{2}}{k_{\|}^{2}} f_{0 e}^{\prime}\left(v_{ph,\parallel}\right) \\
&=-\frac{\pi}{2} \frac{\omega}{k_{\|} v_{T \parallel}} \frac{1}{k_{\|}^{2} \lambda_{D}^{2}} \frac{1}{\sqrt{2}} v_{T \parallel} f_{0 e}\left(v_{ph,\parallel}\right) H_{1}\left(\frac{v_{ph,\parallel}}{\sqrt{2} v_{T \parallel}}\right) .
\end{aligned}
\end{equation}

The overall amplification of beam instability and Landau damping on any initial perturbation wave field $E_1$ can be expressed as $E_1 e^{(\gamma_{t r} + \gamma_{L e}) \Delta t}$. Taking the phase mixing effect into account, the beam distribution is smoothed within a few trapping periods. This timescale can be approximated as the inverse of phase mixing rate of trapped electrons $\Delta t \sim 1 / \gamma_{\text{mixing}}$, which can be estimated from the trapping frequency:
\begin{equation}\label{eq:gamma_mixing}
\frac{\gamma_{\text {mixing }}}{k_{\|} v_{T \|}} \sim \frac{\omega_{t r}}{k_{\|} v_{T \|}} =\frac{k_{\|}^{\prime}}{k_{\|}}\cdot \frac{\omega_{tr}}{k_{\|}^{\prime}v_{T_\parallel}}  =\frac{1}{\kappa}\sqrt{\frac{e \delta \phi}{T_{\|}}},
\end{equation}
where the $\kappa$  represents the wave number ratio of EAW and whistler: $k_\parallel / k_{\parallel} ^{\prime}$. Thus, the critical condition for EAW or nonlinear electrostatic structure excitation can be written as 
\begin{equation}\label{eq:critical-inequality}
    \gamma_{t r}+\gamma_{L e} \geq N \gamma_{\text {mixing }} ,
\end{equation} 
which means that the signal is observable after $N$ $e$-foldings. Combining with Equations \eqref{eq:gamma-tr}--\eqref{eq:critical-inequality}, we obtain the explicit critical condition as a function of $v_{ph,\parallel}/v_{T \parallel}$ and $e\delta \phi / T_{\parallel}$:
\begin{equation}\label{eq:critical-inequality-expansion}
\begin{gathered}
\frac{\sqrt{\pi}}{4} \frac{v_{ph,\|}}{ v_{T_\|}} \exp(-\frac{v_{ph,\|}}{ 2v_{T_\|}})\left[\frac{\pi}{2} \sum_{n=0}^{\infty} \frac{H_{2 n+1}\left(\frac{v_{p h, \|}}{\sqrt{2} v_{T_\|}}\right)}{(2 n+1) !}\left(\frac{e \delta \phi}{2 T_{\|}}\right)^n\right. \\
\left.-H_1\left(\frac{v_{p h, \|}}{\sqrt{2} v_{T_\|}}\right)\right] \geq \frac{k_{\|}^2 \lambda_D^2 N}{\kappa} \sqrt{\frac{e \delta \phi}{T_{\|}}} = {{k_{\|}^{\prime}}^2 \lambda_D^2 N \kappa}\sqrt{\frac{e \delta \phi}{T_{\|}}} = {{k_{\|}^{\prime}}^2 {d_e}^2  \frac{v_{T_\parallel}^2}{c^2} N \kappa}\sqrt{\frac{e \delta \phi}{T_{\|}}}.
\end{gathered}
\end{equation}
\add{where the $d_e$ is the electron inertial length.}

\begin{figure}[htpb]
\noindent\includegraphics[width=\textwidth]{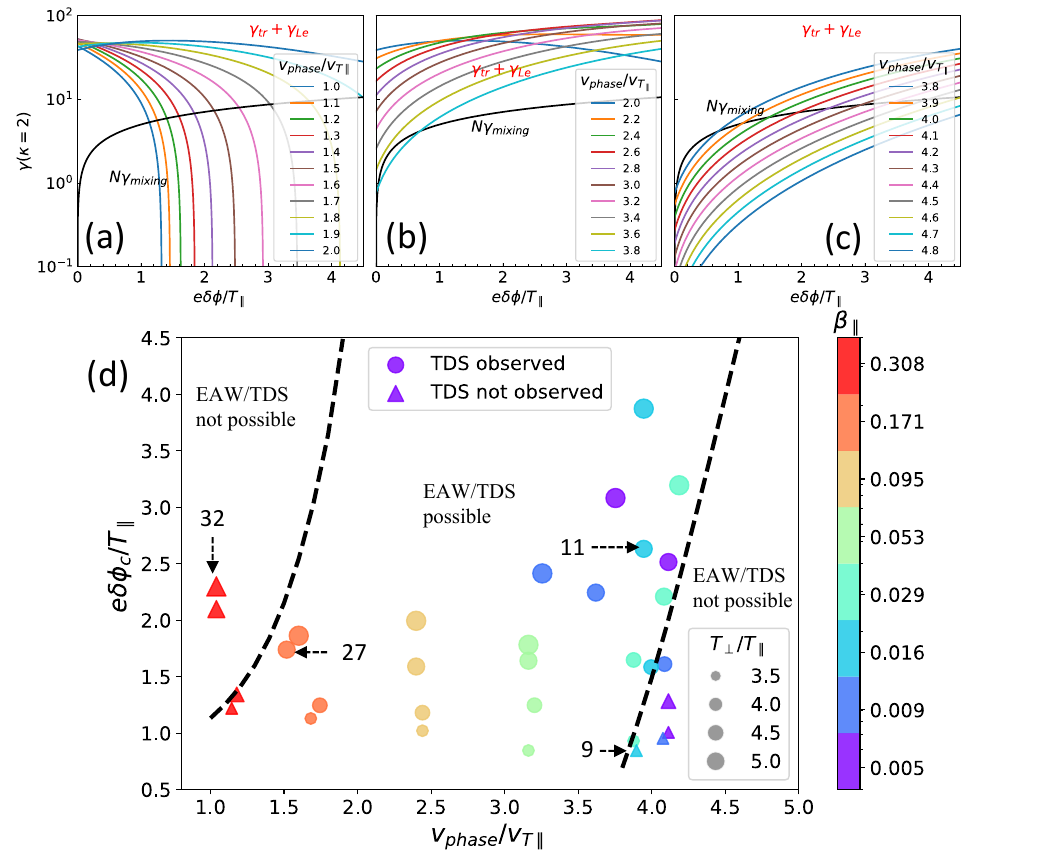}%
\caption{\label{Figure 5} The critical condition for nonlinear excitation of EAWs. (a)-(c) The comparison between the EAW growth rate and the phase mixing rate with different $v_{ph,\parallel} / v_{T\parallel}$. In evaluating the left-hand side of Inequality \eqref{eq:critical-inequality}, we use the typical parameter $k_{\parallel}^{\prime}\lambda_D = 0.05$ ($k_{\parallel}^{\prime}\lambda_D$ being in the range $0.04$--$0.06$ in the simulations), truncate the power series at $n = 4$ (the relative error being $<5 \times 10^{-5}$), and parameterize the growth rate by $v_{ph,\parallel} / v_{T\parallel}$. In evaluating the phase mixing rate on the right-hand side of Inequality \eqref{eq:critical-inequality}, the number of e-foldings is chosen as $N = 5$ (i.e., the signal-to-noise ratio is $e^5$). The result is shown as a black line. (d) The critical amplitude of whistler waves to drive EAWs or TDSs. The left boundary corresponds to the intersection of the mixing rate and the EAW/TDS growth rate in Panel (a), and the right boundary corresponds to that in Panel (c). The critical amplitudes are plotted for $\kappa = 2$. The scattered dots show the PIC simulation results. The circles/triangles are results with/without EAW excitation. We use the average parallel electric field at saturation time in the simulations to estimate $\delta \phi = |\delta E_x| / k^{\prime} $}. 
\end{figure}

We proceed to evaluate both sides of the inequality \eqref{eq:critical-inequality} and display them in three distinct regimes of $v_{ph,\parallel}/v_{T \parallel}$ [Figures \ref{Figure 5}(a), \ref{Figure 5}(b) and \ref{Figure 5}(c)]. We choose typical $k_{\|}^{\prime} \lambda_D = 0.05$, $\kappa = 2$ and N = 5 in our simulation. In Figure \ref{Figure 5}(a), for $1 \leq v_{ph,\parallel}/v_{T \parallel} \leq 1.9$, the EAW growth rate ($\gamma_{tr} + \gamma_{Le}$) drops below the phase mixing rate ($N \gamma_{\text{mixing}}$) at a critical value $\delta \phi_c$ ($1.1 \leq \frac{e \delta \phi_c}{T_{\parallel}} \leq 4.5$). Beyond this value, the excitation of EAWs is suppressed because the trapped electron beam is smoothed before EAWs grow to an observable level. In Figure \ref{Figure 5}(b), for $2 < v_{ph,\parallel}/v_{T\parallel} < 3.6$, the EAW growth rate exceeds the phase mixing rate for any wave potential $\delta \phi$. In Figure \ref{Figure 5}(c), for $3.8 \leq v_{ph,\parallel}/v_{T\parallel} \leq 4.8$, the EAW growth rate exceeds the phase mixing rate at a critical value, above which EAWs can be excited by the trapped electron beam. Thus, we plot the critical amplitude for a wide range of $v_{ph,\parallel}/v_{T\parallel}$. The boundaries given by the critical wave amplitudes divide the parameter space of ($v_{ph,\parallel}, \delta \phi$) into three regions [Figure \ref{Figure 5}(d)]: An excitation band in the middle where the excitation of EAWs is allowed, and two stop bands at the two ends where the excitation of EAWs is prohibited.

\begin{figure}[htpb]
\noindent\includegraphics[width=\textwidth]{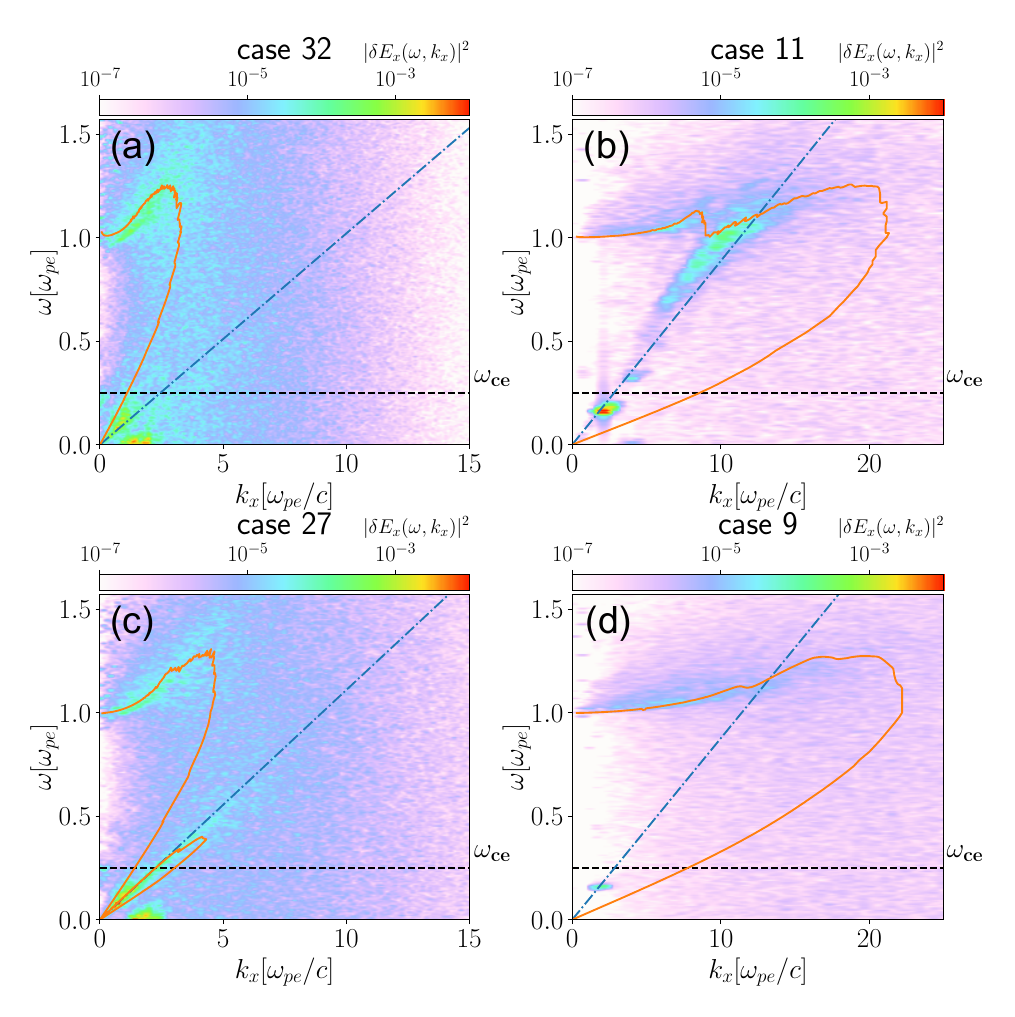}%
\caption{\label{Figure 6} The dispersion diagrams of Fourier-transformed $\delta E_x$ as a function of frequency and wave number in four cases as marked in Figure \ref{Figure 5}(d). The dispersion relations are calculated using the parallel electron distributions after wave saturation. (a) Case $32$. (b) Case $11$. (c) Case $27$. (d) Case $9$. The electron distribution functions used to calculate the dispersion relations are taken at $t = 1200 \omega_{pe}^{-1}$, $t = 1700 \omega_{pe}^{-1}$, $t = 2000 \omega_{pe}^{-1}$, and $t = 4600 \omega_{pe}^{-1}$ in the respective simulations in panels (a)--(d).}%
\end{figure}

We further compare the critical condition for EAW excitation with our PIC simulation results. As shown in Figure \ref{Figure 5}(d), those simulations without EAW excitation are located in the stop band or near the boundaries between the stop and excitation bands, which supports our theoretical estimation. EAWs cannot be excited in cases where $\beta_{\parallel} > 0.3$ and $\beta_{\parallel} < 0.016, T_{\perp}/T_{\parallel} \leq 4$. We choose four representative cases [indicated by arrows and case numbers in Figure \ref{Figure 5}(d)] to illustrate their dispersion diagrams in Figure \ref{Figure 6}. Figures \ref{Figure 6}(a) and \ref{Figure 6}(c) show the two cases (NO.~$32$ and $27$) on two sides of the left boundary \change{in}{from} the ``small" $\beta_{\parallel}$ regime. Although the wave amplitude $\delta \phi$ of case $32$ is larger than that of case $27$, EAWs are excited in the latter case, but not in the former case, consistent with the theoretical model. In comparison, Figures \ref{Figure 6}(b) and \ref{Figure 6}(d) show the two cases (NO.~$11$ and $9$) near the right boundary \change{in}{from} the ``large" $\beta_{\parallel}$ regime. Case $11$ having a wave amplitude larger than the critical amplitude can excite EAWs, whereas case $9$ with a wave amplitude near the critical amplitude cannot excite EAWs. Interestingly, the ``thumb'' dispersion relation of EAWs/Langmuir waves for a Maxwellian distribution\cite{valentini2006excitation} is strongly bifurcated into new branches at the whistler/EAW phase velocities in the two cases with EAWs [cases $27$ and $11$], due to the finite plateau width in the electron distributions (created by whistler waves). The EAWs in cases $11$ and $27$ are mainly different in their phase velocities and range of wave numbers: case $11$ with a high phase velocity is near the Langmuir branch, while case $27$ with a low phase velocity is in the dispersionless acoustic branch. Such differences are governed by the ratio of whistler phase velocity to electron thermal velocity $v_{ph,\parallel}/v_{T_\parallel}$ and the finite resonant island in the electron distribution functions\cite{an2019unified}.

\section{Summary}\label{sec:summary}
In this study, using a series of 2D PIC simulations, we demonstrate the excitation of EAWs and nonlinear electrostatic structures through the nonlinear Landau resonant interaction between whistler waves and electrons. We further derive a critical condition for such excitation of EAWs and nonlinear electrostatic structures in the parameter space of whistler wave amplitude and phase velocity, which shows good agreement with the simulation results. The main results are as follows. 
\begin{enumerate}
\item In all our PIC simulations, whistler waves are naturally generated through the temperature anisotropy instability. In the small $\beta_{\parallel}$ regime of $\beta_{\parallel}\lesssim 0.025$, oblique, quasi-electrostatic whistler waves are excited, whereas in the large $\beta_{\parallel}$ regime of $\beta_{\parallel}\gtrsim 0.025$, quasi-parallel, electromagnetic whistler waves are excited. In both regimes, whistler waves can have strong enough parallel electric fields to accelerate/form electron beams in their potential wells, i.e., nonlinear Landau resonance.

\item Electron beams trapped by whistler waves subsequently excite EAWs, which may further evolve into TDSs nonlinear electrostatic structures (time domain structures, TDS). The EAW phase velocities are approximately equal to the beam velocities, or equivalently the whistler phase velocities. We obtain the dispersion relation of EAWs using the electron distributions from PIC simulations, and show that the finite plateau distribution (created by the beam) allows the survival of EAWs even when their phase velocities are close to the electron thermal velocity.

\item We derive the critical condition for EAW excitation by comparing the EAW growth rate and the phase mixing rate of a trapped electron beam. This critical condition is constructed in the parameter space of normalized whistler wave amplitude $e \delta \phi / T_\parallel$ and normalized whistler phase velocity $v_{ph,\parallel} / v_{T \parallel}$: At $v_{ph,\parallel} / v_{T \parallel} \lesssim 2$, there exists an upper bound of $e \delta \phi / T_\parallel$ for EAW excitation; At $v_{ph,\parallel} / v_{T \parallel} \gtrsim 3.6$, there exists a lower bound of $e \delta \phi / T_\parallel$ for EAW excitation; At $2 \lesssim v_{ph,\parallel} / v_{T \parallel} \lesssim 3.6$, EAWs are unconditionally excited. These theoretical predictions are consistent with the PIC simulation results.
\end{enumerate}

The modulation of high-frequency electrostatic waves (either Langmuir or electron acoustic waves) by whistler waves has been widely observed in Earth's inner magnetosphere\cite{li2017chorus}, magnetotail\cite{chen2022high}, magnetopause reconnection region\cite{li2018local,wang2023electrostatic}, and planetary magnetospheres\cite{reinleitner1984chorus}. Our results provide a clear, quantitative explanation for these observations. Particularly, the critical condition for EAW excitation can be tested against in-situ spacecraft observations. Moreover, such coupling from whistler to high-frequency electrostatic waves indicates a channel of energy transfer across different spatial scales. Taking the fast plasma injections from Earth's magnetotail to the inner magnetosphere as an example, whistler waves, which are generated by mesoscale electron injections, transfer their energy to the high-frequency, Debye-scale electrostatic waves. The coupling process involving nonlinear Landau resonance serves as a cross-scale energy channel for the dissipation of injected energy in the form of electron heating \cite{vasko2017diffusive,vasko2015thermal,osmane2014threshold,artemyev2014thermal,an2021nonlinear}.

\section*{Data Availability}

The data that support the findings of this study are available from the corresponding author upon reasonable request.


%
%

%

\begin{acknowledgments}
This work was supported by NASA awards 80NSSC20K0917, 80NSSC22K1634, and 80NSSC23K0413, NSF award 2108582, and NASA contract NAS5-02099. We would like to acknowledge high-performance computing support from Cheyenne (doi:10.5065/D6RX99HX) provided by NCAR's Computational and Information Systems Laboratory, sponsored by the National Science Foundation \cite{cheyenne}. We would also like to thank E. Paulo Alves for insightful discussions, and the OSIRIS Consortium, consisting of UCLA and IST (Lisbon, Portugal) for the use of \texttt{OSIRIS} and for providing access to the \texttt{OSIRIS} 4.0 framework.
\end{acknowledgments}

\bibliographystyle{elsarticle-harv}
\bibliography{eaw_whistler}

\end{document}